\documentclass[12pt]{article}

\usepackage{epsfig}
\usepackage{graphicx}
\usepackage{color}
\usepackage{bm}

\newcommand{\simg}{\stackrel{\sim}{>}}

\begin{document}

\title{Numerical study of a disordered model for DNA denaturation 
transition}
\author{Barbara Coluzzi}

\maketitle

\begin{center}
{\em Service de Physique Th\'eorique} - CEA - Saclay - Orme des Merisiers 91191 Gif-sur-Yvette cedex, France

\begin{abstract}
\noindent
We numerically study a disordered version of the model for DNA denaturation 
transition (DSAW-DNA) consisting of two interacting SAWs in $3d$, which 
undergoes a first order transition in the homogeneous case. 
The two possible values ${\hat \epsilon_{AT}}$ and ${\hat \epsilon_{GC}}$ 
of the interactions between base pairs are taken as quenched random variables 
distributed with equal probability along the chain. We measure quantities
averaged over disorder such as the energy density, the specific heat
and the probability distribution of the loop lengths. When applying the 
scaling laws used in the homogeneous case we find that
the transition seems to be smoother in the presence of disorder,
in agreement with general theoretical arguments, although
we can not rule out the possibility of a first order transition.
\end{abstract}

\end{center}


\section{Introduction}
\noindent
The DNA denaturation transition is an old-standing open problem \cite{review} 
and one finds in the literature a large number of different models that 
look in detail at various aspects \cite{PoSc}-\cite{IvZeZo}, 
for instance 
reproducing the double-helix structure and the corresponding denaturation 
behaviour in the whole temperature-torsion plane \cite{CoMo} or taking into 
account to different extents the role of the stacking energies 
\cite{DaPeBi,IvZeZo}. 
From the experimental point of view, one observes a multi-step 
behaviour in light absorption as a function of temperature which suggests 
a sudden sharp opening of clusters of base pairs in cooperatively melting 
regions. Therefore one expects that a theoretical model, correctly 
reproducing the experimental behaviour, should undergo a sharp transition and 
there has recently been a lot of interest in models 
possibly displaying 
a first order transition in the homogeneous case \cite{CoMo}-\cite{RiGu}.

Most programs, for example MELTSIM \cite{Bl}, predict the 
experimentally observed melting curves using a model originally introduced 
by Poland and Scheraga \cite{PoSc,RiGu} that 
takes into account the different entropic weights of opened loops 
and double stranded segments. In the first studies of this model, excluded
volume effects were completely neglected and a smooth second order
transition was predicted in two and three dimensions. 
By solving the model including the entropic weights of self-avoiding 
loops \cite{Fi}, one finds a sharper but still second order 
transition. In this way the self-avoidance between bases within 
the same loop is taken into account, but other mutual excluded 
volume effects are still neglected. 

In a previous work \cite{CaCoGr} we pointed out the importance of 
excluded volume effects between different segments and loops
by introducing and numerically studying a model inspired by the former
of Poland and Scheraga, consisting of two interacting self-avoiding walks 
on a $3d$ lattice (SAW-DNA).

In the limit of infinite chain length, the SAW-DNA model undergoes a first 
order phase transition from the double strand to the molten single-stranded 
chain state when varying the temperature. The order parameter, which is 
the energy or the density of binded base pairs, varies abruptly from 
one to zero.

It has been theoretically demonstrated 
that the transition in the homogeneous model by Poland and Scheraga 
is of first order when excluded volume effects are completely taken into
account, using the entropic weight of a self-avoiding loop embedded in
a self-avoiding chain \cite{KaMuPe1,KaMuPe2}. This
can be obtained as a particular case of a more
general result on polymer networks \cite{Du}. It has
also been analytically shown that only considering the
self-avoidance between the two different chains leads 
to a first order transition \cite{GaMoOr,BaCaOrSt}. 

Moreover, several numerical investigations \cite{CaOrSt,BaCaOrSt,
BaCaSt,BaCaKaMuOrSt} 
of the homogeneous SAW-DNA model carefully measured 
the probability distribution of the lengths of denatured loops $P(l)$ 
at the critical point in $3d$ and $2d$. These confirmed the theoretically 
expected power law $P(l) \propto 1/l^c$ with exponent $c>2$ (in 
agreement with the transition being first order) in both dimensions.

In the same SAW-DNA model, a first order unzipping transition is predicted in 
the presence of an external force that pulls apart the two DNA strands 
\cite{KaMuPe2}. The behaviour in the temperature-force plane is numerically 
investigated in \cite{OrBhMaMaSe}, and a phase diagram with a re-entrance 
region is observed.

The version of the Poland-Scheraga model on which programs
for predicting experimental denaturation curves are based, 
contains another main 
parameter, the cooperativity factor $\sigma$, which takes into account 
the activation barrier to open a loop and has the effect of 
sharpening the transition when considering finite size chains. 
It was shown \cite{BlCa} that by using the value $c \simeq 2.15$ that 
characterizes $P(l)$ in $3d$ (instead of the usual value $c \simeq 1.75$, 
i.e. the 
exponent of an isolated self-avoiding loop), and by slightly
correspondingly varying the cooperativity factor, one still obtains 
a multi-step melting behaviour well in agreement with experimental 
results. Nevertheless, the relevance of self-avoidance for the 
experimental DNA denaturation is still an open question \cite{HaMe,GaOr}. 

In this work, we present numerical results on the denaturation 
transition in the
interacting SAW model in $3d$ in the presence of quenched disorder (DSAW-DNA).
The main aim of our study is to attempt to clarify whether
disorder is relevant, therefore we introduce a model as simple as 
possible, since we expect that slightly more realistic versions should behave 
similarly.
Studies in the literature on the effects of sequence heterogeneities on 
other simple models for DNA denaturation \cite{CuHwa,LuNe} neglect 
self-avoidance. Apart from its importance to experimental
melting, we find that this is an intriguing
statistical mechanics model in itself. On the one hand, from general 
theoretical arguments \cite{Ha}-\cite{Be}, one may expect that the
transition should become smoother in the presence of disorder, 
although this is a peculiar kind of first order phase transition, 
corresponding to a tri-critical point in the fugacity-temperature plane,
 and it is characterized both by an $\alpha=1$ specific
heat exponent and by a diverging correlation length \cite{CaCoGr}. 
On the other hand, numerical results on $P(l)$ 
at the critical 
temperature for a single
disordered sequence in this model \cite{CaOrSt} seem to show that the
order of the transition does not change, as also suggested
from the topological considerations explaining the sharp transition 
in the homogeneous case. 

We shall consider the simple case in which there are only two 
possible base pair interactions, i.e. ${\hat \epsilon_{AT}}$ 
which describes the Adenine-Thymine couple and ${\hat \epsilon_{GC}}$ for the
Guanine-Cytosine one, with ${\hat \epsilon_{GC}}=2{\hat \epsilon_{AT}}$. 
To be realistic, one should choose closer values but,
if disorder is relevant, they could make it difficult to find numerical 
evidence for its effect, being reasonable to expect that the disorder 
possibly changes 
the order of the phase transition 
as soon as  ${\hat \epsilon_{AT}} \neq  {\hat \epsilon_{GC}}$. 
Therefore we assumed to take ${\hat \epsilon_{GC}}=
2{\hat \epsilon_{AT}}$ as an interesting compromise.

Moreover, we take the interactions to be independent quenched random 
variables, identically distributed with equal probability 1/2. 
Again, it should be noted that 
in more realistic models the interaction energies are chosen to depend also on 
the first neighbours (to take into account at least partially the stacking 
energies) and that there are probably long-range correlations in the 
base pair distribution, the ratio of the GC to AT content being in any event 
a highly varying quantity usually far from 1. We are neglecting both of these
effects, but we find that our simplified model does already display an 
intriguingly rich behaviour. It seems to be a useful starting point 
for understanding the thermodynamical properties of this kind of systems in 
the presence of disorder and their relevance for describing experimental DNA 
denaturation transitions.

\section{Model and Observables}
\noindent
Let us define two $N$-step chains with the same origin on 
the 3-$d$ lattice, $\omega^1 = \{\omega^1_0, \dots, \omega^1_N\}$ and 
$\omega^2 =\{\omega^2_0, \dots, \omega^2_N\}$, with $\omega^k_i \in Z^3$ 
and $\omega^1_0=\omega^2_0 = (0,0,0)$.

The Boltzmann weight of a configuration $(\omega^1,\omega^2)$ of our system is
\begin{equation}
\exp \left ( {-H \over k_B T} \right ) = \prod_{i\ne j}(1-\delta_{\omega^1_i,\omega^2_j})
(1-\delta_{\omega^1_i,\omega^1_j})(1-\delta_{\omega^2_i,\omega^2_j})
\exp\left(-\sum_{i=0}^N { \epsilon_i \over k_B T} \delta_{\omega^1_i,\omega^2_i}\right),
\end{equation}
where the $\{ \epsilon_i \}$ are independent quenched random variables,
identically distributed according to the bimodal probability
\begin{equation}
{\cal P}(\epsilon)=\frac{1}{2} \left [ \delta ( \epsilon - {\hat \epsilon_{AT}} )
+ \delta ( \epsilon - {\hat \epsilon_{GC}} ) \right ].
\end{equation}

Thermodynamic properties of the system only depend on the
reduced variables ${\hat \epsilon_{AT}}/k_B T$ and 
${\cal R}={\hat \epsilon_{AT}}/{\hat \epsilon_{GC}}$, the homogeneous case
corresponding to ${\cal R}=1$. Here we will take ${\hat 
\epsilon_{AT}=1}$ and ${\hat \epsilon_{GC}=2}$, i.e. ${\cal R}=1/2$.
For the sake of clarity, we also fix the Boltzmann constant to $k_B=1$, 
measuring 
the temperature in ${\hat \epsilon_{AT}}$ units.

Let us introduce the different observables in the well studied homogeneous 
case. In the thermodynamical limit, the transition is a tri-critical
point in the fugacity-temperature ($z-T$) plane \cite{CaCoGr}, described
by the crossover exponent $\phi$. One has:

\begin{equation}
z-z_c \propto (T-T_c)^{\phi}
\end{equation}

The order parameter characterizing the transition is the density of closed
base pairs, that behaves like the energy density $e=E_N/N$, where 
\cite{CaCoGr,EiKrBi}: 
\begin{equation}
E_N(T) \sim \left\{
\begin{array}{ccc}
1/(T-T_c) & \hspace{.3in} & T \rightarrow {T_c}^+ \\
N^{\phi} & \hspace{.3in} & T=T_c \\
{N} ({ T_c}-{T})^{1/{\phi}-1} & \hspace{.3in} & T \rightarrow {T_c}^- 
\end{array}
\right .
\end{equation}
Therefore the value $\phi=1$ of the crossover exponent corresponds to
a first order transition, in which $e$ goes
discontinuously (in the thermodynamical limit) from the zero value of
the high-temperature coiled phase to a finite value at $T_c$. We stress
again that it is a peculiar kind of first order transition with
a diverging correlation length and absence
of surface tension (the probability distribution of the energy
is nearly flat at the critical temperature).

In the homogeneous case, both when self-avoidance is completely taken
into account (i.e., a first order transition with $\phi=1$ in $d=3$ and
$d=2$) and when it is neglected (the random walk model, which undergoes 
a second order transition with $\phi=1/2$ in $d=3$ and a first order 
transition for $d \ge 5$), one finds that the finite size behaviour
of different quantities is in agreement with given scaling laws.
The total energy $E_N$, its probability distribution $P(E_N)$, and
the maximum of the specific heat $c^{max}_N$ behave as 
\cite{CaCoGr,EiKrBi}:

\begin{eqnarray}
E_N(T)/N^{\phi} & = & {\tilde h}[(T-T_c)N^{\phi}] 
\label{enelaw} \\
c^{max}(N) & \propto & N^{2\phi-1} \label{claw} \\
P_N(E)N^{\phi} & = & {\tilde f}(E/N^{\phi}) \hspace{.3in} {\mbox{ at} }
\hspace{.3in} T=T_c, 
\label{plaw}
\end{eqnarray}
where ${\tilde f}$ and ${\tilde h}$ are scaling functions.

Moreover, one can get an independent evaluation of the crossover exponent
from the probability distribution of the loop lengths, that at 
$T_c$ is in agreement with the law:
\begin{equation}
P(l) \propto \frac{1}{l^c},
\label{plclaw}
\end{equation}
where the exponent is related to $\phi$ through the equation 
$\phi = \min \{ 1,c-1 \}$. In the $3d$ SAW-DNA,
one finds numerically \cite{BaCaSt,BaCaKaMuOrSt} $c=2.14(4)$, in perfect
agreement with the theoretical prediction \cite{KaMuPe1} $c \simeq 2.115$, 
that implies a first order transition. 

It is also interesting to note
that $P(l,T)$ is more generally expected to behave as \cite{BaCaKaMuOrSt}:
\begin{equation}
P(l,T) \propto {\exp(-l/\xi(T)) \over l^c},
\label{pgenlaw}
\end{equation}
and that the correlation length $\xi(T)$ should diverge when approaching the
transition point with the power law:
\begin{equation}
\xi(T) \propto \left\{
\begin{array}{lcl}
| T-T_c |^{-1} & \hspace{.3in} & \mbox{ for } c \ge 2 \nonumber \\
| T - T_c|^{-1/(c-1)} & \hspace{.3in} & \mbox{ for } 1 < c \le 2. \\
\end{array}
\right.
\end{equation}

In the presence of the independent quenched random variables $\epsilon=
\{ \epsilon_i \}$,
one has to introduce quantities averaged over disorder:
\begin{eqnarray}
E_N(T) & = & \overline{E_{N,\epsilon}(T)} \\
c_{N,T} & = & \overline{c_{N,\epsilon}(T)} \\
P_N(E) & = & \overline{P_{N,\epsilon}(E)} \\
P_N(l,T) & = & \overline{P_{N,\epsilon}(l,T)},
\end{eqnarray}
where
\begin{equation}
\overline{O(T)} = \int d\epsilon {\cal P}(\epsilon) O_{\epsilon}(T), 
\end{equation}
the numerical evaluation of these quantities being as usually performed by 
averaging over a (large) number of different disordered configurations 
(samples). One should note that the energy density 
$e=-n_{AT} \hat{\epsilon}_{AT}-n_{GC}\hat{\epsilon}_{GC}$ is now 
quantitatively different from (minus) the number of contact density 
$n=n_{AT}+n_{GC}$, though it seems reasonable to expect that the two
quantities behave similarly.

In particular, we will look at the previously discussed scaling laws 
(\ref{enelaw})-(\ref{pgenlaw}) on the 
averaged energy, the averaged specific heat maximum, the averaged 
probability distribution of the energy and the averaged probability 
distribution of the loop lengths. 

\section{Simulations}
\noindent
We used the pruned-enriched Rosenbluth method (PERM) \cite{Gr}, with
Markovian anticipation \cite{FrCaGr}, which is particularly effective
to simulate interacting polymers \cite{CaCaGrPe}. Moreover, we used an  
$ad~hoc$ bias for the present model. 
When the second chain has to perform a growth step and the end of the first 
one is in a neighboring site, instead of doing a blind step multiplying 
the weight by a factor $e^{-{\epsilon_i}/{k_B T}}$ if the new contact is 
formed,
we favor contacts choosing the step towards the end of the first chain with
an appropriately higher probability and correcting the weight accordingly.

For each considered sequence we have performed 16 independent runs at
different temperature values with $3 \div 5$ million
independent starts for each run. This turns out to generate enough 
statistics up to chain lengths $N=800$, since the number of times that
the system reaches the largest $N$ value in the simulation is of the 
order of the number of independent starts. 

We compute the behaviour of the energy, of its probability distribution
and of its derivative (i.e., the specific heat) as a function of
temperature by re-weighing the data at the chosen temperatures
of the set. The statistical errors on the values for a given sequence are 
 evaluated using the Jack-knife method and we checked that they are 
definitely smaller than the errors due to sample-to-sample fluctuations, 
which are our estimate of the errors on averaged quantities.

Moreover, we checked thermalization and the correct evaluation of the
errors, particularly in the case of $P(l,T)$, by comparing results 
obtained at two different sets of temperatures (see the Table)
with two different methods. In the first case two copies of the systems 
evolved simultaneously and independently (the algorithm being
accordingly modified) and the computed quantities are practically
evaluated from two independent runs of 3 millions of starts each. In the
second set one copy performed 5 millions of independent starts. We obtained
perfectly compatible results from the two sets of simulations.

We considered 128 different samples. We note that the
disorder configurations for different $N$ values are not
completely independent, nevertheless it seems reasonable to expect
to observe the same scaling properties.
Therefore we find our statistics to
be sufficient for giving a first insight onto the behaviour of various 
quantities
and we also stress that the crossover exponent can in principle 
be obtained from data on $P_N(l)$ at $T_c$
corresponding to a single $N$ value. 
The whole simulation took
about 15000 hours (on COMPAQ SC270 and, to a lesser extent, on 
Forshungszentrum J\"ulich CRAY-T3E).

\begin{table}
\begin{center}
\begin{tabular}{||l||c|c|c|c|c|c|c|c||}
\hline
\hline
$T^{a}_i$ & 0.8 & 0.95 & 1.08 & 1.125 & 1.15 & 1.175 & 1.2 & 1.3 \nonumber \\
\hline
\hline
$T^{b}_i$ & 0.875 & 1.015 & 1.05 & 1.1 & 1.13 & 1.16 & 1.19 & 1.25 \nonumber \\
\hline
\hline
\end{tabular}
\caption{The considered temperatures for the two sets of simulations
performed for each sample. In the first case (a) two copies of the
system evolved independently and statistics were collected over 3 millions 
of starts, whereas in the second case (b) only one copy was simulated 
and statistics were collected over 5 millions of 
starts.}
\end{center}
\label{temperatures}
\end{table}

\section{Results}
\noindent
We shall focus on results and compare them to what is observed
in the ordered case.

\begin{figure}[hpbt]
\epsfig{file=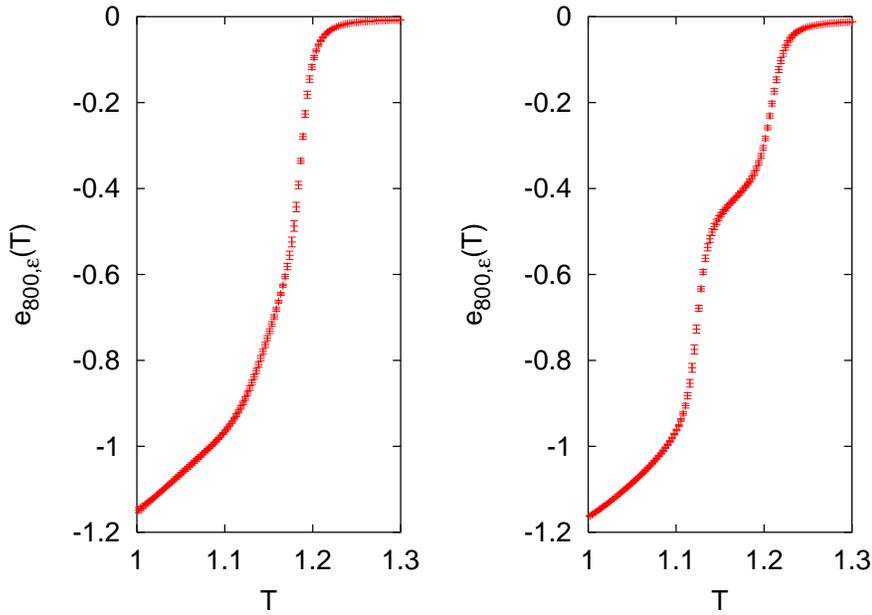,angle=270,width=12cm}
\caption{
The energy density $e_{N,\epsilon}(T)$ for two particular 
sequences with $N=800$, one of which displays a two-step behaviour. 
Note the plateau at $e_{N,\epsilon}(T)=0$ in the coiled phase. The plotted
region is the one around $T_c$ in which the energy varies more rapidly 
and there are evident differences in the behaviour from sample to sample.
At lower temperatures quantities are nearly sample-independent and the energy
density slowly decreases towards the fully double-stranded limit 
${\overline{e_{min,\epsilon}}}=-1.5.$}
\label{ene_sample}
\end{figure}

\begin{figure}[hpbt]
\epsfig{file=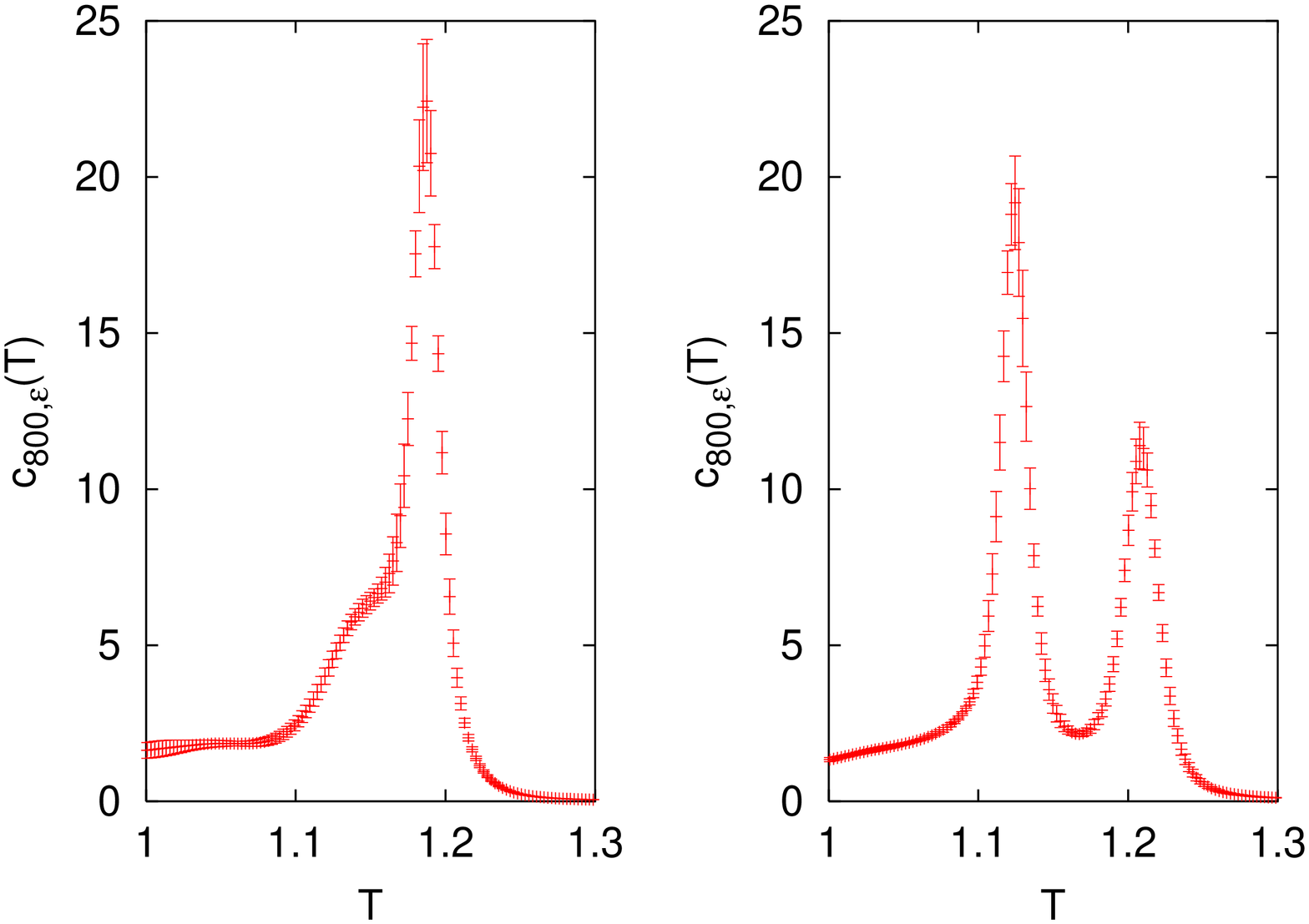,angle=270,width=12cm}
\caption{
The specific heat $c_{N,\epsilon}(T)$ for two different disordered 
sequences with $N=800$ (the same sequences as in previous figure). 
Also in
this case there is qualitative agreement with the experimentally observed
behaviour in the differential melting curves, though the model should be 
improved in order of really comparing. For instance, particularly for short 
sequences, the plateau one gets in the coiled phase is usually higher than the 
low temperature one because of residual stacking energies which is an effect 
completely neglected here.} 
\label{cv_sample}
\end{figure}

First of all we find, 
at least for some sequences (about one third of the sequences for the
largest length), 
the expected multi-step behaviour of the energy density 
(see Fig.\ \ref{ene_sample}) and correspondingly 
several peaks of the specific heat (see Fig.\ \ref{cv_sample}), 
that behaves qualitatively as the derivative of the 
density of closed base pairs with respect to the temperature, 
i.e. the differential melting curve which is usually experimentally 
measured. 

Qualitatively speaking, the presence of multi-steps becomes
more probable with increasing $N$, whereas the temperature region in which
this behaviour is present becomes narrower. This is in agreement with the 
observation that at the critical point
the usual argument for proving the self-averaging property of 
the densities of extensive quantities, such as the energy density, 
fails since one can not divide the system in nearly 
independent sub-systems 
due to the diverging correlation length. This is true also in the 
SAW-DNA where, despite 
the first order transition of the homogeneous case, one finds a diverging
correlation length. On the 
considered $N$ range we observe a strong sample-dependent behaviour, as in 
the case also of $P_{N,\epsilon}(l)$ at $T \sim T_c$. It is therefore possible
that the results of the application of the scaling law of the homogeneous
case to averaged quantities have to be taken with some care \cite{WiDo}.

\begin{figure}[hpbt]
\epsfig{file=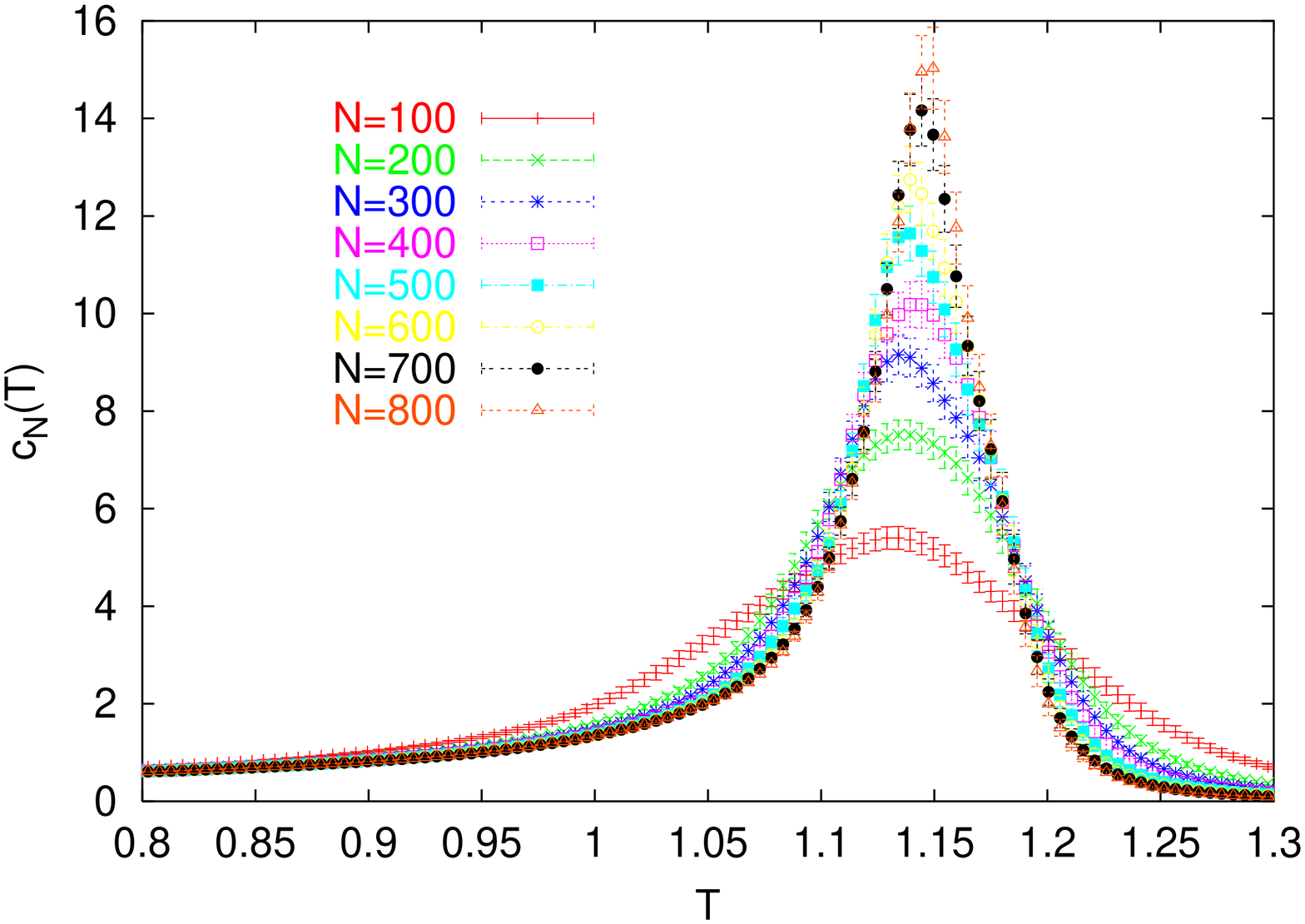,angle=270,width=12cm}
\caption{ 
The disorder averaged specific heat for the considered
chain lengths.\label{calspec}}
\end{figure}

In Fig.\ \ref{calspec} we plot our data on the disorder averaged specific
heat for the considered chain lengths. This shows that,
also in the presence of disorder, the maximum of the specific heat  
 increases as a function of the chain length. Nevertheless,
as discussed in detail in the following, 
it seems to behave as $N^{2\phi-1}$ with exponent
smaller than one.

\begin{figure}[hpbt]
\epsfig{file=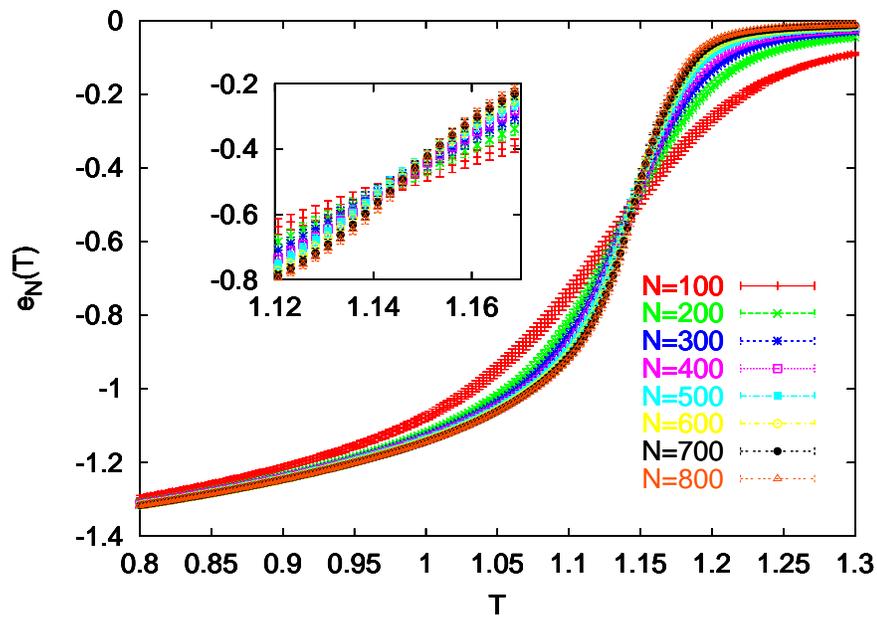,angle=270,width=12cm}
\caption{ 
The disorder averaged energy density, for the considered
chain lengths.\label{ene}}
\end{figure}

\begin{figure}[hpbt]
\epsfig{file=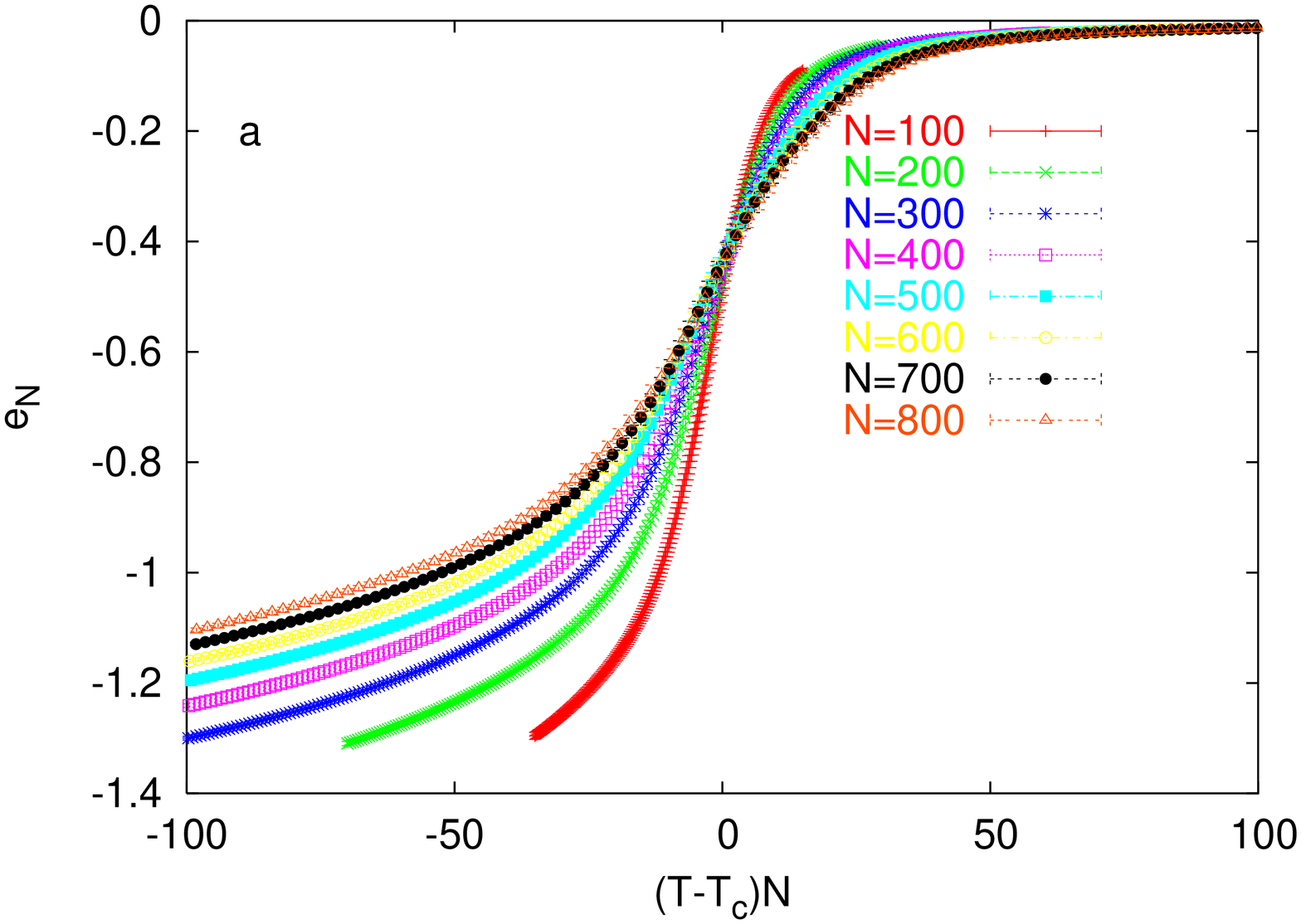,angle=270,width=12cm}
\epsfig{file=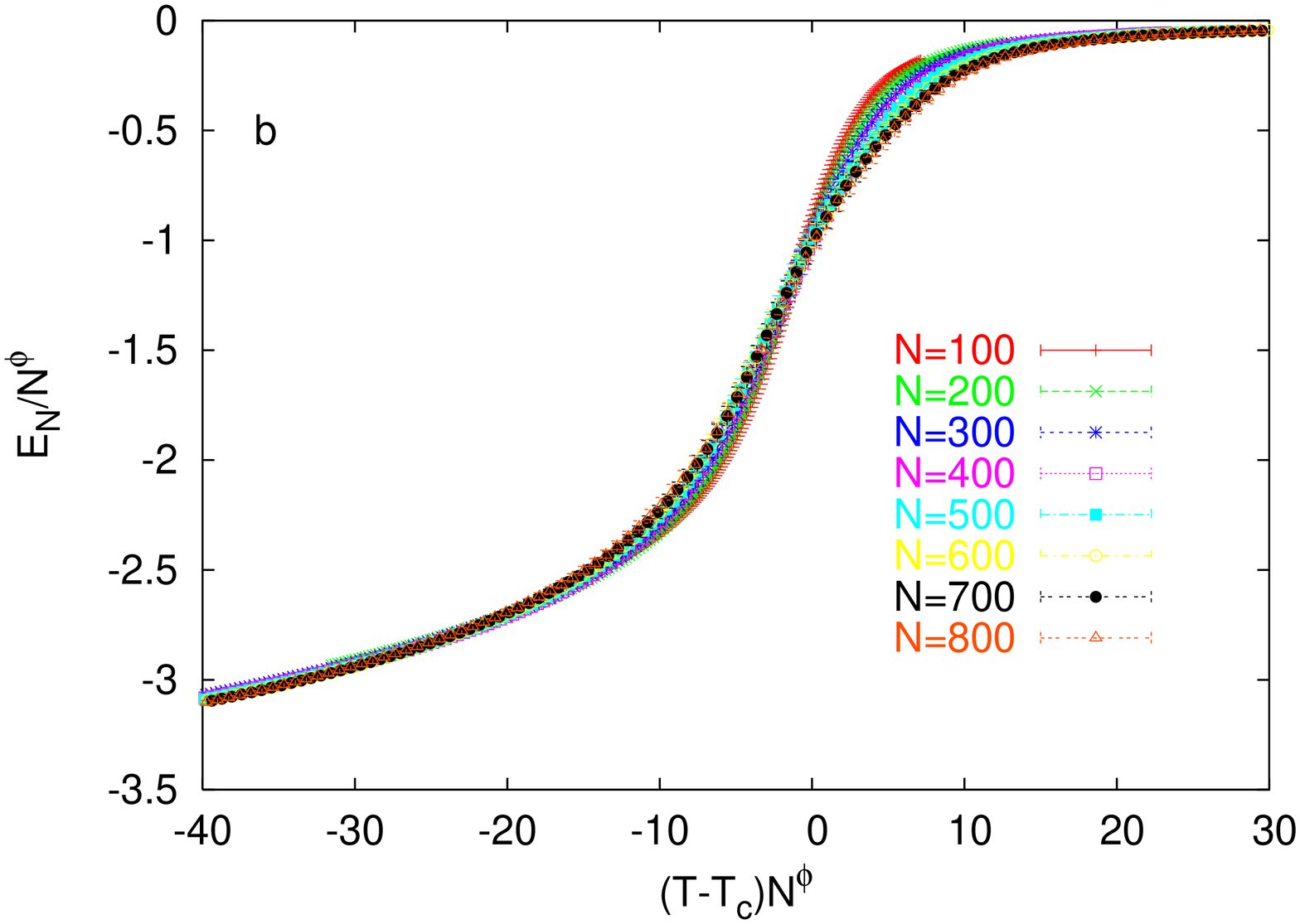,angle=270,width=12cm}
\caption{ 
On the top we plot the disorder averaged energy density for the 
considered 
chain lengths as a function of $N(T-T_c)$ with $T_c=1.15$. On the bottom 
we present the scaling law with $\phi= 0.8475$ and 
$T_c = 1.155$.\label{enescala}}
\end{figure}

Then in Fig.\ \ref{ene} we present our data on the disorder averaged energy 
density.
We note that curves corresponding to different $N$ values cross
roughly at the same temperature $T_c \simeq 1.15$ within the errors 
(see the insert), therefore suggesting a transition still of
first order. Actually this should mean a jump of the
energy density in the thermodynamical limit, from zero for $T>T_c$ 
to the finite value $e(T_c) \simeq -0.5$. Nevertheless, as displayed 
in Fig.\ \ref{enescala}a, the expected scaling law Eq.\ (\ref{enelaw}),
which is well verified in the homogeneous case at least 
in the region $T \simg T_c$,
is here definitely not fulfilled with $\phi=1$. When fitting
the data according to this scaling law (see Fig.\ \ref{enescala}b) 
one finds a slightly higher
critical temperature value $T_c = 1.155$ and a crossover
exponent $\phi = 0.8475$. 

Since we were looking for a second order transition, i.e.
for an exponent $\phi < 1$, we required 
in the fit that data obey the scaling law on both sides of the
critical temperature (for the random walk model
in $3d$ \cite{CaCoGr}, where the transition is of second order,  a good 
scaling was observed on both sides of $T_c$).
In any event we stress that by fitting data only in the high-temperature
region $T \simg T_c$ one would get a definitely lower value of the crossover
exponent $\phi \sim 0.6$. 

We note that the data roughly agree with the law but there are still 
corrections to scaling. We also checked that the disorder averaged
number of contact density displays a similar behaviour (i.e. it follows
the same scaling law).

\begin{figure}[hpbt]
\epsfig{file=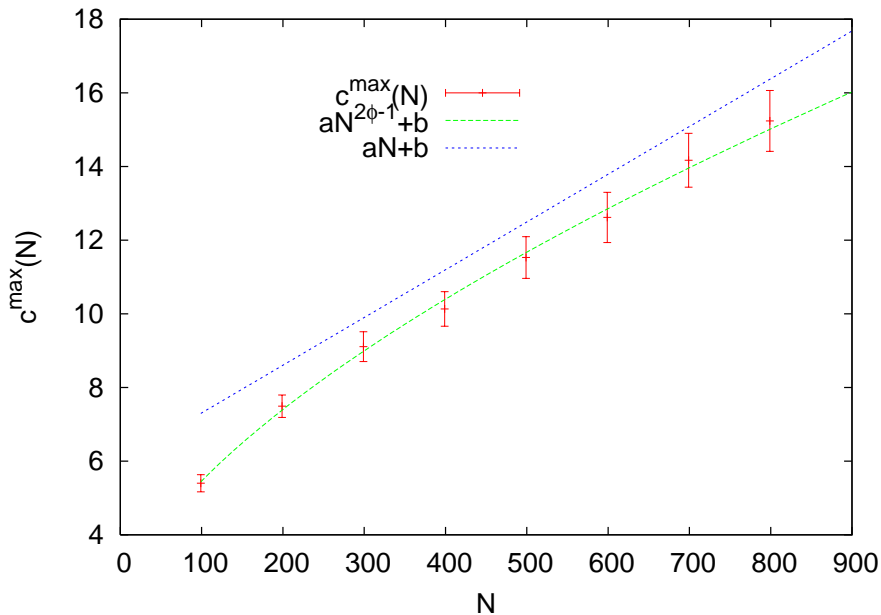,angle=270,width=12cm}
\caption{ The behaviour of the specific heat maximum compared
with a fit to the exponent $\phi=0.815$ and to $\phi=1$.
\label{calspecmax}}
\end{figure}

In Fig.\ \ref{calspecmax} we present our data on the specific heat maximum, 
that would agree well with an exponent $\phi \simeq 0.85$ (our fit to the
expected behaviour Eq. (\ref{claw}) gives $\phi = 0.815$).
However the data would also be compatible (within the errors) 
with $\phi=1$, if we were to neglect the smallest chain length
$N=100$. 

We tried, following \cite{WiDo}, to analyze the data on the energy densities 
and the specific heats in terms of a reduced 
sample dependent temperature $T-T^i_c(N)$, where $T^i_c(N)$ is defined as
the temperature for which the specific heat for sample $i$ and chain length
$N$ reaches its maximum. This analysis leads to the same conclusion
as the conventional one, that 
the disorder averaged energy density does not 
scale with a $\phi=1$ crossover exponent and the disorder averaged 
specific heat maximum
seems to diverge more slowly than linearly with $N$ when taking into account
all the considered chain lengths. We computed the average distance
between the sample dependent critical temperatures 
$dT_c(N) = \frac{1}{{\cal N}({\cal N}-1)}
\sum_{i \neq j} |{T_c}^i(N)-{T_c}^j(N)|$ (${\cal N}=128$ being the number of 
samples) which seems to go to zero for increasing
chain lengths more slowly than $1/N$, 
again suggesting a second order
transition. 

\begin{figure}[hpbt]
\epsfig{file=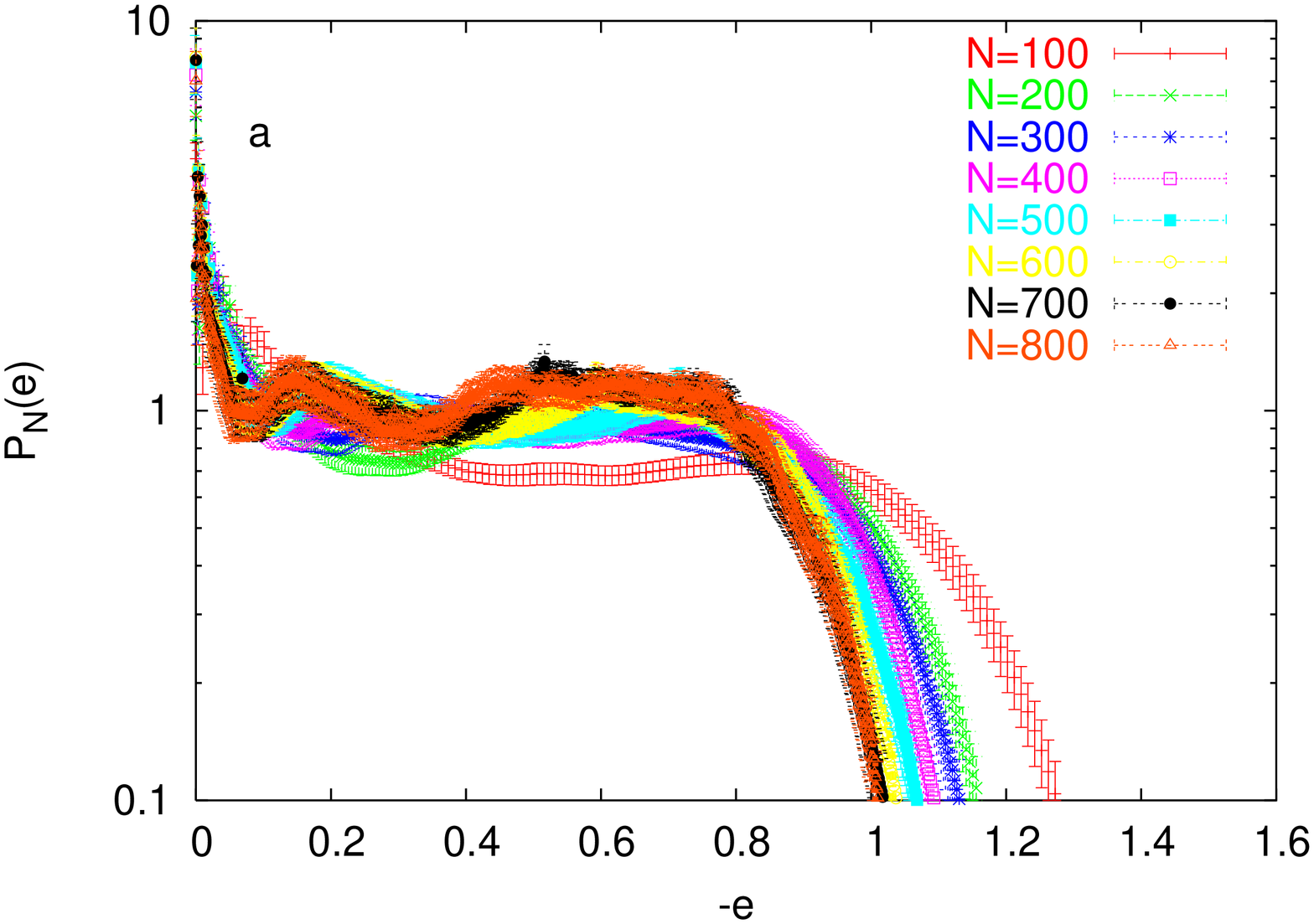,angle=270,width=12cm}
\epsfig{file=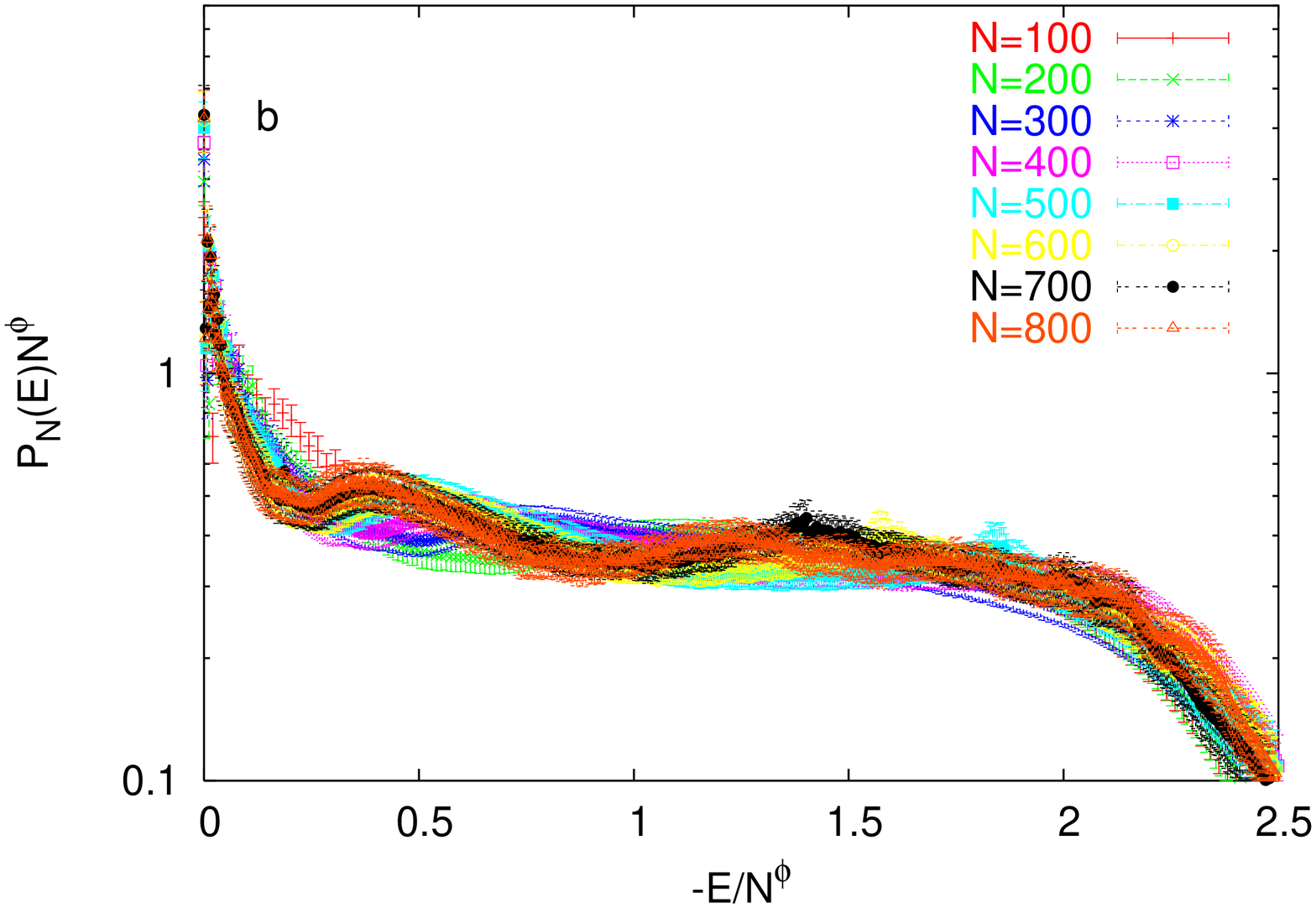,angle=270,width=12cm}
\caption{ 
On the top we plot the disorder averaged probability distribution 
of the energy density at $T=1.15\simeq T_c$, 
$P_N(e)$ as a function of $-e$, which
corresponds to the particular case of the corresponding scaling law with 
$\phi=1$, since $P_N(e)=N\:P_N(E/N)$. 
On the bottom we present the scaling law,  
$P_N(E)N^{\phi}$ as a  
function of $-E/N^{\phi}$, at $T_c=1.155$ with $\phi=0.8475$.}
\label{pe}
\end{figure}

We looked at the probability distribution of the energy density
 at the critical temperature $T = 1.15 
\simeq T_c$ 
(see Fig.\ \ref{pe}a).
In the homogeneous case it displays a large and flat scaling region extending
to a value $e^*$ which do not depend on $N$, with deviations from
the scaling law Eq.\ (\ref{plaw}) for $e < e^*$ \cite{CaCoGr}.
Here we find that the scaling law with $\phi=1$ is not well fulfilled.
 For the sake of comparison we also plot (in Fig.\ \ref{pe}b) the behaviour 
of our data at the slightly higher
temperature value $T=1.155$ with a crossover exponent
$\phi = 0.8475$.

\begin{figure}[hpbt]
\epsfig{file=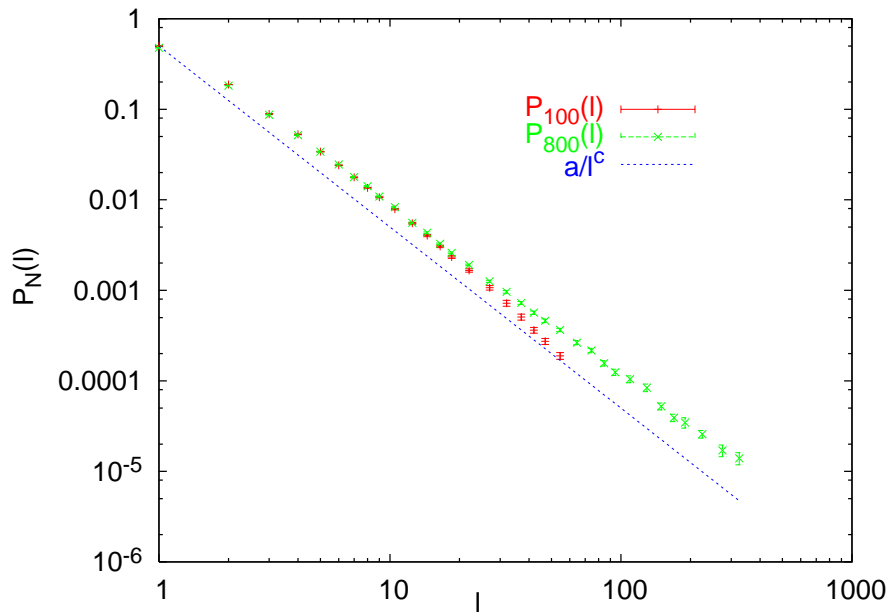,angle=270,width=12cm}
\caption{ 
The disorder averaged probability distribution of the loop lengths 
at $T=1.15 \simeq T_c$ for
$N=100$ and $N=800$, compared to the behaviour $a/l^c$ with $c=2$.\label{pl}}
\end{figure}

Let us now consider data on the disorder 
averaged probability distribution of the
loop lengths ${P_N(l)}$ which are plotted in Fig.\ \ref{pl} at
$T=1.15 \simeq T_c$. We checked that the behaviour is very similar
when  temperature slightly varies. 
Again, a fit to the expected power law
$\propto 1/l^c$ of the whole data set suggests an exponent
$c < 2$, i.e. a second order transition. Nevertheless, a closer inspection
of the behaviour makes evident some curvature and when restricting the
fit to the $l$ range $1<<l<<N$ (which is also the region in that
the hypothesis $P_N(l) \propto 1/l^c$ should be better verified) 
we find larger $c$ values and the data for $N=100$ alone are 
definitely consistent with a first order transition.

\begin{figure}[hpbt]
\epsfig{file=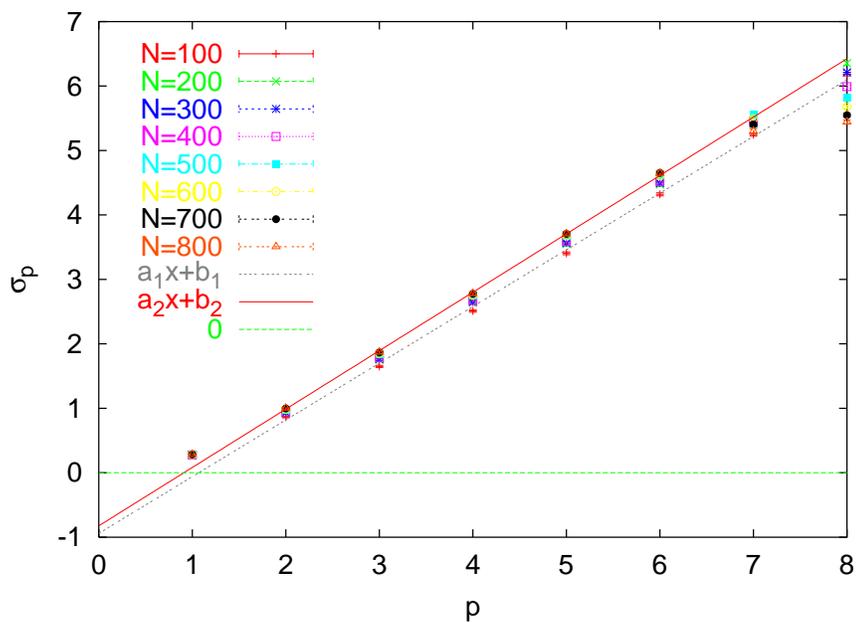,angle=270,width=12cm}
\caption{ 
Data on $\sigma_p =  \ln  ({\overline{<l^p>}}) / \ln N$, from data on 
$P(l)$ at $T=1.15 \simeq T_c$, plotted as 
function
of $p$. The two fits give $c = 2.07$ and $c = 1.91$ respectively. We note that 
the linear behaviour is no more satisfied for $p>6$ in data corresponding to 
the largest $N$ values.\label{sigma}}
\end{figure}

 To perform a more quantitative
analysis, in Fig.\ \ref{sigma} 
we consider \cite{TeMeSt} the momenta ${\overline{<l^p>}}$, which
are expected to behave as $N^{\sigma_p}$. In particular the combination 
$\ln {\overline{<l^p>}}/ \ln N$ is expected to be linear in $p$ on a large 
range of $p$ values and one should be able to evaluate $c-1$ by extrapolating 
the linear
behaviour down to $p=0$. We effectively observe a quite linear behaviour
for $p>1$ but by extrapolating we get the definitely different values 
$c = 2.07$ for $N=100$ and $c=1.91$ for $N=800$. On the one hand
data for the longest chain length should be the most meaningful, therefore
confirming a second order transition; on the other hand it should
be stressed that $P_{N,\epsilon}(l)$ is a difficult quantity to measure and 
we can not rule out the possibility that our statistics are inadequate for 
correctly evaluating it for the largest considered chain lengths.

Nevertheless, the observed curvature on the $\log P_N(l,T_c) \: vs
\log(l)$ behaviour and the corresponding strong $N$-dependence in the
obtained estimations could be related to finite size effects. One should
note in particular that in the case of a second order transition the finite
size corrections to the critical temperature (i.e. $T_c(N)-T_c(\infty)$) 
are expected to approach zero in the thermodynamical limit more slowly than 
$1/N$. Therefore, it could be misleading to fit the data according to the
power law $P(l)\propto 1/l^c$ which is expected to be fulfilled only 
exactly at $T_c$. Moreover we have a large uncertainty on the evaluation 
of $T_c$ itself. For these reasons, in order to gain further insight into the 
behaviour of the system, we also attempt to compare the measures at the
different temperatures considered with the more general law 
$P_N(l,T) \propto \exp(-l/\xi_N(T))/l^c$. This should allow to take partially
into account finite size effects, by introducing a possible finite
correlations length $\xi_N(T)$ which measures both the distance from the
effective $T_c(N)$ and the finite size corrections to the 
thermodynamical limit correlation length. 

Interestingly enough, in the temperature region $T \simg T_c \sim 1.15 $
one gets $1/\xi_N(T)$ compatible with zero within the errors (it can be
also negative, though small) apart from the shortest chain 
lengths. In particular, data for $N=100$ are consistent with a detectably
finite correlation length for all the temperatures studied.
The corresponding evaluation of the exponent $c$ is in this
case smaller than that found when the effect of the finite correlation
length is neglected, giving usually $c<2$. 

In detail, we performed three-parameter fits of data on $P_N(l,T)$ at different 
temperatures by neglecting both the very first and the 
last $l$ values. In particular the dependence on the number of neglected 
initial points has 
been studied by restricting the range to $l>5$ and to $l>2$. 
Whenever the fit gives a $1/\xi_N(T)<0$ unphysical value, the 
corresponding $c$ is estimated from the power law behaviour, i.e. by imposing 
$1/\xi_N(T)=0$. Moreover we also considered the fits to the power law 
in all the cases in which an inverse correlation length compatible with
zero turns out, again by looking 
only at the $l>5$ range or at the whole $l>2$ one.  
Finally, we performed linear extrapolations to the $1/N \rightarrow 0$ limit.

Despite the introduction of the correlation length, the evaluations
of $c$ seem generally to depend on temperature and on the number
of initial points included in the fit, 
and its extrapolations turn out to be compatible with a first order
transition. Nevertheless, the values obtained in the region $T \simeq T_c$, 
where the largest sizes can be fitted to the power law and which are 
therefore expected to be the most significant results, 
would definitively suggest a second order transition, characterized by a quite 
large crossover exponent $\phi \sim 0.9$. It should be noted that in this
region one gets well compatible estimations from the different methods
we used. 
In conclusion, the analysis is consistent 
with the initial qualitative observation on the $P(l)$ behaviour at 
$T \simeq T_c$, that the transition seems to be of second order when 
considering the longest chain lengths most meaningful.

It would be interesting to perform on $P_N(l,T)$ an analysis by
introducing rescaled temperature but we unfortunately are not able to
re-weigh data at different $T$ values in this case. To study whether
this would not affect the qualitative behaviour, thereby confirming
a second order transition, is left for future investigation.

\begin{figure}[hpbt]
\epsfig{file=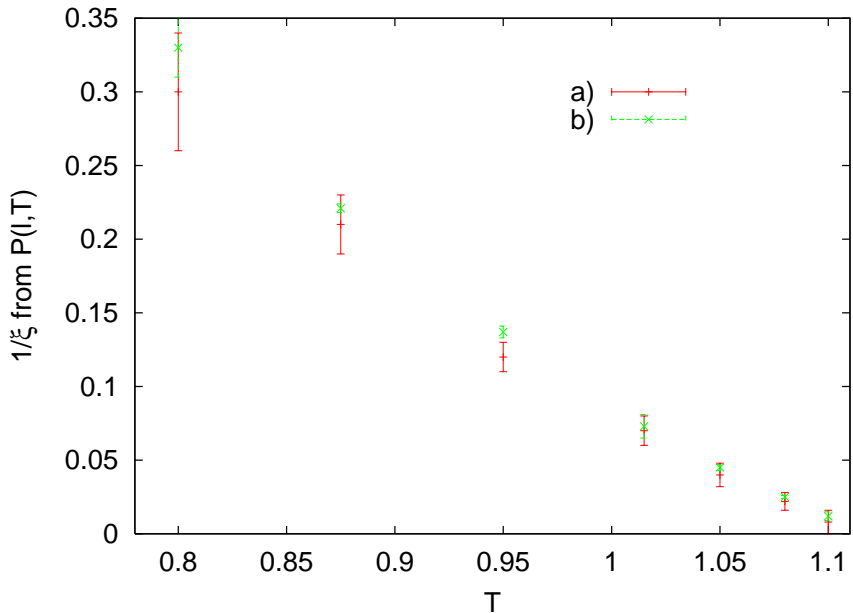,angle=270,width=12cm}
\caption{ 
The results on the inverse correlation length $1/\xi(T)$ behaviour of the 
fits of data on $P_N(l,T)$ as obtained by different
methods: a) For each $N$ and temperature the values are obtained from a 
three-parameter fit to the law $\propto \exp({-l/\xi_N(T))
/l^c}$ neglecting only the very first (in particular the first two) and 
the last $l$ values. b) The same as in a) but restricting the considered range
to $l>5$. In both cases values for different $N$ are 
linearly extrapolated to $1/N \rightarrow 0$ in order to give the plotted 
estimations. We stress that because of the fits involved the given errors are 
to be considered only very indicative, our uncertainty on the estimations 
being better expressed by the fluctuations between values obtained with the 
different methods.}
\label{chi_fit}
\end{figure}

Finally, we present in Fig. \ref{chi_fit} data on the inverse 
correlation length, $1/\xi(T)$ as evaluated by extrapolating linearly to the 
limit $1/N \rightarrow 0$ values for the different chain lengths (here we only 
consider the temperature range where it is definitely larger than zero). 
The obtained qualitative behaviour suggests again a second order transition, 
since $1/\xi(T)$ seems to go to zero more rapidly than $(T_c-T)$. 
A fit to the expected power law behaviour 
$1/\xi(T) \propto (T_c-T)^{1/(c-1)}$ would give
$c \sim 1.7$. We checked (on the $N=800$ data) that the qualitative behaviour
of $\xi(T)$ does not change if we were to fit the data on $P_N(l,T)$ in the low
temperature region according to the law (\ref{pgenlaw}) by imposing $c=2$ (or
$c=2.1$). 

As a last remark, we stress that our statistics are unfortunately
inadequate for performing a more quantitative analysis, particularly
on $\xi(T)$. Actually, because of the algorithm we are using,
data on $P(l,T)$ in the low temperature range are reliable up to smaller
$l$ values rather than in the region $T \simg T_c$ (i.e. very long bubbles are
usually not well sampled at low temperatures due to their negligible
probability). Once again, we are led to the conclusion that the most
significant results on $P(l,T)$ should be those obtained in the $T \simeq T_c$
region for the largest chain lengths, which is our best numerical evidence
for the transition being of second order.

We have 
some preliminary numerical results on our model with different 
values of ${\cal R}$, where the crossing of the energy densities becomes less 
evident the more different  
the energies ${\epsilon_{AT}}$ and ${\epsilon_{GC}}$ are.
The transition seems to be characterized by a varying 
crossover exponent which becomes smaller as ${\cal R}$ diminishes. 
In particular, by applying the scaling law to the disorder averaged energy 
densities we get $\phi \sim 0.8$ from data for ${\epsilon_{AT}}=1$ and 
${\epsilon_{GC}}=4$, and $\phi \sim 0.7$ from data for ${\epsilon_{AT}}=0$
and ${\epsilon_{GC}}=1$. The $P(l)$ behaviour at the critical temperature
would suggest
similar values too, though also in these cases one gets different results
by restricting the range to the region $1<<l<<N$.  
Here we only would like to mention
that an alternative explanation for these results is 
that one is always looking at a first order transition but that the scaling 
laws of the homogeneous case are not well fulfilled anymore \cite{WiDo}. More
extensive simulations would be necessary in order to clarify this issue.

\section{Conclusions}
Summarizing, we studied a disordered model for DNA denaturation transition 
consisting of two interacting self-avoiding chains in which we chose 
the two possible values of the interaction energy to be ${\epsilon}_{GC}=
2{\epsilon}_{AT}$, distributed with equal probability. Despite of the 
extensive numerical simulations performed, it is difficult to definitively 
discriminate 
between a first order (as in the homogeneous case)
and a smoother second order transition. 

As a matter of fact, we find that the 
energy densities as a function of temperature for the different chain lengths 
considered roughly cross within the errors therefore suggesting a 
discontinuity of this quantity in the thermodynamical limit. However,
the application of the scaling laws 
that are verified in the ordered case indicate a smoother 
second order 
transition with strong corrections to scaling.

When looking at the averaged probability distribution of the loop lengths,
one gets again a crossover exponent $\phi$ definitely smaller than one from
the data corresponding to the largest considered chain lengths. Nevertheless, 
data display some curvature 
also at $T \simeq T_c$ and higher values
of the exponent are obtained when restricting the range to $1 << l <<N$.
A momenta analysis of this probability distribution gives $c<2$
(i.e. $\phi<1$) only
for the largest lengths. 
For a better understanding of this issue, we also attempted to perform a more 
general analysis on the whole considered temperature range by introducing a 
finite correlation length, with the aim of partially
taking into account finite size effects. Results in the region $T \simeq T_c$
point towards a $c<2$ value and are our best numerical evidence
of a second order transition. Anyway, it should be stressed that we get a quite 
large $c \sim 1.9$
and we can not rule out the possibility of a sharper (first order)
transition.

During completion of this work, results appeared on the homogeneous 
Poland-Scheraga model
\cite{Sc} that seem to show how this model with the appropriate parameter
values and the lattice SAW-DNA model are 
equivalent, even though the lattice model displays strong finite size 
corrections to scaling.
Therefore a different kind of analysis on the disordered Poland-Scheraga
model could help in better understanding the situation.
To this extent, the recent study by Garel and Monthus \cite{GaMo} 
points towards
the direction of a first order transition also in the presence of disorder,
though the usual scaling laws seem to be not fulfilled. On the other hand,
a very recent theoretical work \cite{GiTo} predicts that disorder should
be relevant in this kind of models and that one should find a second order
(or smoother) transition. In a nutshell, the situation appears far from being
clarified and these simple models for DNA denaturation transition
seem to deserve a careful analysis from the statistical mechanics point
of view. 

Finally, it should be pointed out
 that even if the transition is smoother in the presence of 
disorder the obtained value  
$c \sim 1.9$ for the considered 
interaction energies is  
definitely higher than the value 
$c=1.76276$ \cite{BeNi} which one gets in the homogeneous case 
when self-avoidance 
between different loops and segments is neglected. In the hypothesis
that $c$ depends on the energy values one would expect, in the 
more realistic case
of ${\epsilon}_{GC}$ closer to ${\epsilon}_{AT}$,
an higher value of $c$ possibly indistinguishable from the
case of a first order transition with $c>2$. 
In this
sense the model seems relevant for the experimentally observed 
DNA denaturation, and
which value of $c$ one should use in predictions based on Poland-Scheraga 
models as realistic as possible seems to be an open question.

\vspace{.5cm}

\section*{Acknowledgments} 
\noindent
It is a pleasure to thank Serena Causo and
Peter Grassberger, who were involved in the beginning of this study. I also
acknowledge stimulating discussions with Enrico Carlon, 
David Mukamel, Henri Orland, Enzo Orlandini, 
Attilio Stella and Edouard Yeramian. I am moreover indebted to Alain Billoire 
for
carefully reading the manuscript and to Thomas Garel and
Cecile Monthus for suggestions and discussion of their
results on the disordered Poland-Scheraga model with $c=2.15$ \cite{GaMo}.
This work was partly funded by a Marie Curie (EC) fellowship 
(contract HPMF-CT-2001-01504).


\begin{thebibliography}{99}

\bibitem{review}
R.M. Wartell and A.S. Benight, {\em Phys. Rep.} {\bf 126}, 67 (1985).

\bibitem{PoSc}
D. Poland and H.A. Scheraga,  {\em J. Chem. Phys.} {\bf 45}, 1456 (1966);
{\bf 45}, 1464 (1966). For a review see D. Poland and H.A. Scheraga (eds.),
{\em Theory of Helix-Coil Transitions in Biopolymers}, (Academic,
New York, 1970). 

\bibitem{DaPeBi} 
M. Peyrard and A.R. Bishop, {\em Phys. Rev. Lett.} {\bf 62}, 2755 (1989);
T. Dauxois, M. Peyrard and A.R. Bishop {\em Phys Rev. E} {\bf 47}, 684 (1993).
S. Ares $el~al$, {\em Phys. Rev. Lett.} {\bf 94}, 035504 (2005).

\bibitem{CoMo} 
S. Cocco and R. Monasson,
{\em Phys. Rev. Lett.} {\bf 83}, 5178 (1999).

\bibitem{IvZeZo}
V. Ivanov, Y. Zeng and G. Zocchi, {\em Phys. Rev. E} {\bf 70}, 051907 (2004);
V. Ivanov, D. Piontkovski and G. Zocchi, {\em Phys. Rev. E} {\bf 71}, 041909 
(2005).

\bibitem{ThDaPe}
N. Theodorakopoulos, T. Dauxois and M. Peyrard,
{\em Phys. Rev. Lett.} {\bf 85}, 6 (2000).

\bibitem{CaCoGr}
M.S. Causo, B. Coluzzi, and P. Grassberger,
{\em Phys. Rev. E} {\bf 62}, 3958 (2000).

\bibitem{KaMuPe1}
Y. Kafri, D. Mukamel, and L. Peliti,
{\em Phys. Rev. Lett.} {\bf 85}, 4988 (2000).

\bibitem{CaOrSt}
E. Carlon, E. Orlandini, and A.L. Stella,
{\em Phys. Rev. Lett.} {\bf 88}, 198101 (2002).

\bibitem{GaMoOr}
T. Garel, C. Monthus, and H. Orland,
{\em Europhys. Lett.} {\bf 55}, 132 (2001);
S.M. Bhattacharjee, {\em Europhys. Lett.} {\bf 57},
772 (2002); T. Garel, C. Monthus, and H. Orland,
{\em Europhys. Lett.} {\bf 57}, 774 (2002).

\bibitem{RiGu} For a recent review in which the effects of self-avoidance
in Poland-Scheraga models are critically discussed see C. Richard and
A. J. Guttman, {\em J. Stat. Phys.} {\bf 115}, 943 (2004).

\bibitem{Bl} R. D. Blake et al, {\em Bioinformatics},
{\bf 15}, 370 (1999).

\bibitem{Fi}
M.E. Fisher,  {\em J. Chem. Phys.} {\bf 45}, 1469 (1966).

\bibitem{KaMuPe2}
Y. Kafri, D. Mukamel, and L. Peliti,
{\em Eur. Phys. J. B} {\bf 27}, 135 (2002).

\bibitem{Du}
B. Duplantier, {\em Phys. Rev. Lett.} {\bf 57}, 941 (1986);
{\em J. Stat. Phys.} {\bf 54}, 581 (1989).

\bibitem{BaCaOrSt}
M. Baiesi, E. Carlon, E. Orlandini, and A. L. Stella,
{\em Eur. Phys. J. B} {\bf 29}, 129 (2002).

\bibitem{BaCaSt} 
M. Baiesi, E. Carlon, and A. L. Stella,
{\em Phys. Rev. E} {\bf 66}, 021804 (2002); 

\bibitem{BaCaKaMuOrSt}
M. Baiesi, E. Carlon, Y. Kafri, D. Mukamel, E. Orlandini, and A. L. Stella,
{\em Phys. Rev. E} {\bf 67}, 021911 (2003).

\bibitem{OrBhMaMaSe}
E. Orlandini, S. M. Bhattacharjee, D. Marenduzzo, A. Maritan, and F. Seno,
{\em J. Phys. A} {\bf 34}, L751 (2001).

\bibitem{BlCa}
R. Blossey and E. Carlon, 
{\em Phys. Rev. E} {68}, 061911 (2003).

\bibitem{HaMe}
A. Hanke and R. Metzler,  {\em Phys. Rev. Lett.} {\bf 90}, 159801 (2003); 
Y. Kafri, D. Mukamel, and L. Peliti,
{\em Phys. Rev. Lett.} {\bf 90} 159802 (2003).

\bibitem{GaOr}
T. Garel and H. Orland,
{\em Biopolymers} {\bf 75}, 453 (2004).

\bibitem{CuHwa}
D. Cule and T. Hwa,
{\em Phys. Rev. Lett.} {\bf 79}, 2375 (1997).

\bibitem{LuNe}
D.K. Lubensky and D.R. Nelson,
{\em Phys. Rev. Lett.} {\bf 85}, 1572 (2000);
{\em Phys. Rev. E} {\bf 65}, 031917 (2002).

\bibitem{Ha}
A. B. Harris,
{\em J. Phys. C} {\bf 7}, 1671 (1974).

\bibitem{AiWe} M. Aizenman and J. Wehr, {\em Phys. Rev. Lett.} {\bf 62},
2503 (1989); {\em Phys. Rev. Lett.} {\bf 64}, 1311(E) (1990).

\bibitem{Be}
A. N. Berker,
{\em Physica A}{\bf 194}, 72 (1993).
 
\bibitem{EiKrBi}
E. Eisenriegler, K. Kremer, and K. Binder,
{\em J. Chem. Phys.} {\bf 77}, 6296 (1982).

\bibitem{Gr} 
P. Grassberger, {\em Phys. Rev. E} {\bf 56}, 3682 (1997).

\bibitem{FrCaGr}
H. Frauenkron, M.S. Causo, and P. Grassberger, 
{\em Phys. Rev. E} {\bf 59}, R16 (1999).

\bibitem{CaCaGrPe}
S. Caracciolo, M.S. Causo, P. Grassberger, and A. Pelissetto, 
{\em J. Phys. A} {\bf 32}, 2931 (1999).

\bibitem{BeNi}
P. Belohorec and B. Nickel,
{\em Accurate Universal and Two-parameter Model Results
from a Monte-Carlo Renormalization Group Study}, Preprint
- University of Guelph (1997).

\bibitem{WiDo}
S. Wiseman and E. Domany,
{\em Phys. Rev. E} {\bf 52}, 3469 (1995); {\em Phys. Rev. Lett.}
{\bf 81}, 22 (1998); {\em Phys. Rev. E} {\bf 58}, 2938 (1998).

\bibitem{TeMeSt}
C. Tebaldi, M. De Menech, and A.L. Stella,
{\em Phys. Rev. Lett.} {\bf 83}, 3952 (1999).

\bibitem{Sc}
L. Sch\"afer,
{\em Can Finite Size Effects in the Poland-Scheraga Model
Explain Simulations of a Simple Model for DNA Denaturation?},
cond-mat/0502668

\bibitem{GaMo}
T. Garel and C. Monthus, 
{\em J. Stat. Mech.} P06004 (2005).

\bibitem{GiTo} 
G. Giacomin and F.L. Toninelli,
{\em Smoothening effect of quenched disorder
on polymer depinning transition}, math.PR/0506431.

\end{thebibliography}
\end{document}